# Modeling Multi-Hop Semantic Paths for Recommendation in Heterogeneous Information Networks

Hongye Zheng
The Chinese University of Hong Kong
Hong Kong, China

Yue Xing
University of Pennsylvania
Pennsylvania, USA

Lipeng Zhu
Johns Hopkins University
Baltimore, USA

Xu Han
Brown University
Providence, USA

Junliang Du
Shanghai Jiao Tong University
Shanghai, China

Wanyu Cui*
University of Southern California
Los Angeles, USA

*Abstract-This study focuses on the problem of path modeling in heterogeneous information networks and proposes a multi-hop path-aware recommendation framework. The method centers on multi-hop paths composed of various types of entities and relations. It models user preferences through three stages: path selection, semantic representation, and attention-based fusion. In the path selection stage, a path filtering mechanism is introduced to remove redundant and noisy information. In the representation learning stage, a sequential modeling structure is used to jointly encode entities and relations, preserving the semantic dependencies within paths. In the fusion stage, an attention mechanism assigns different weights to each path to generate a global user interest representation. Experiments conducted on real-world datasets such as Amazon-Book show that the proposed method significantly outperforms existing recommendation models across multiple evaluation metrics, including HR@10, Recall@10, and Precision@10. The results confirm the effectiveness of multi-hop paths in capturing high-order interaction semantics and demonstrate the expressive modeling capabilities of the framework in heterogeneous recommendation scenarios. This method provides both theoretical and practical value by integrating structural information modeling in heterogeneous networks with recommendation algorithm design. It offers a more expressive and flexible paradigm for learning user preferences in complex data environments.*

## I. INTRODUCTION

Heterogeneous information networks (HINs) have attracted widespread attention in recent years due to their rich semantic expressiveness and structural representation capabilities in recommendation system research. Unlike traditional homogeneous networks, HINs can represent multiple types of nodes and edges within a unified structure, allowing them to capture complex multi-source heterogeneous relationships. In user-item interactions, there are often multiple types of interaction paths, such as "user-click-item-belong-to-category" or "user-follow-user-purchase-item." These multi-hop paths contain semantic information that provides a more detailed and diverse foundation for recommendation tasks [1]. However, how to effectively identify, model, and utilize the high-order semantics embedded in these multi-hop paths remains a major challenge in current recommendation research.

Multi-hop paths in HINs exhibit complex compositional semantics and contextual relevance. They can reflect users' potential interest preferences and behavioral patterns. Traditional recommendation methods often overlook the structured semantics embedded in these paths and rely only on shallow interactions or first-order neighbor information, which limits their ability to capture users' latent interest transitions and behavioral motivations. Although some existing path-based recommendation methods introduce the modeling of path structures, most rely on manually defined path patterns. They lack the ability to perceive the diversity of path compositions and the dynamic nature of semantics. Therefore, designing a recommendation model with the capability to understand path semantics and adapt to multi-hop information propagation is one of the key directions for enhancing recommendation performance [2].

In addition, not all paths in a heterogeneous information network are equally important. Different paths vary in their ability to describe user interests. For example, some paths may reflect only surface-level interests, while others reveal deeper content preferences. Thus, modeling based solely on predefined rules or fixed path lengths may introduce redundant information or even noise, which could negatively affect the recommendation performance [3]. To address this issue, it is necessary to introduce a path-aware mechanism that enables the model to automatically identify and focus on high-quality paths relevant to the target task. By incorporating appropriate attention mechanisms or path selection strategies, the model's representational power and generalization ability can be significantly improved.

When designing multi-hop path-aware mechanisms, it is also important to consider the interaction and fusion between path contexts. Multi-hop paths exhibit both structural dependency and semantic transitivity. During network propagation, the state of each node in a path may be influenced by the previous node, and the overall semantics of the path may evolve dynamically [4]. Therefore, modeling the high-order dependencies among nodes within a path, the semantic differences between paths, and the overall semantic impact on user interest representation is essential for implementing effective multi-hop path-based recommendation algorithms. Furthermore, path-aware recommendation models should be scalable and adaptive to different types of heterogeneous network structures and recommendation scenarios.

This study focuses on multi-hop paths in heterogeneous information networks and explores their modeling and optimization in recommendation systems. It aims to construct a recommendation framework with both semantic understanding and path awareness. At the theoretical level, this work contributes to a deeper understanding of semantic propagation patterns in heterogeneous structures. At the application level, it addresses current bottlenecks in complex user behavior modeling, multi-source data fusion, and high-order preference capture. It also promotes the advancement of personalized recommendations in domains such as e-commerce, social networks, and content platforms. By integrating multi-hop path modeling mechanisms with semantic-aware approaches, this research is expected to provide a new modeling paradigm and performance improvement path for recommendation algorithms in heterogeneous information networks [5].

## II. METHOD

The multi-hop path-aware recommendation method proposed in this study builds upon the structural and semantic modeling strengths of heterogeneous information networks. The architecture is organized into three interconnected components: path construction and screening, path semantic representation learning, and path-aware fusion prediction, as illustrated in Figure 1. In the first phase, inspired by Liang [6], who demonstrated that path-level attention mechanisms significantly mitigate the impact of data sparsity in user-item interaction graphs, we implement a discriminative path filtering mechanism. This component filters out semantically weak or noisy paths, thus ensuring that the remaining paths carry stronger relational semantics and contribute meaningfully to user preference modeling. The second component—semantic representation learning— researchers proposed adaptive fusion techniques for cold-start scenarios [7]; accordingly, we employ a sequential modeling strategy that simultaneously encodes both entities and relations within a path. This enables the preservation of fine-grained semantic dependencies and contextual information across hops. Finally, in the fusion prediction stage, we incorporate an attention-based weighting strategy to integrate information from multiple semantic paths. This design is partly motivated by Li et al. [8], whose contrastive learning framework highlights the value of differential attention across complex relational structures. By assigning varying importance to different paths, the model generates a refined global user interest representation that drives the final recommendation.

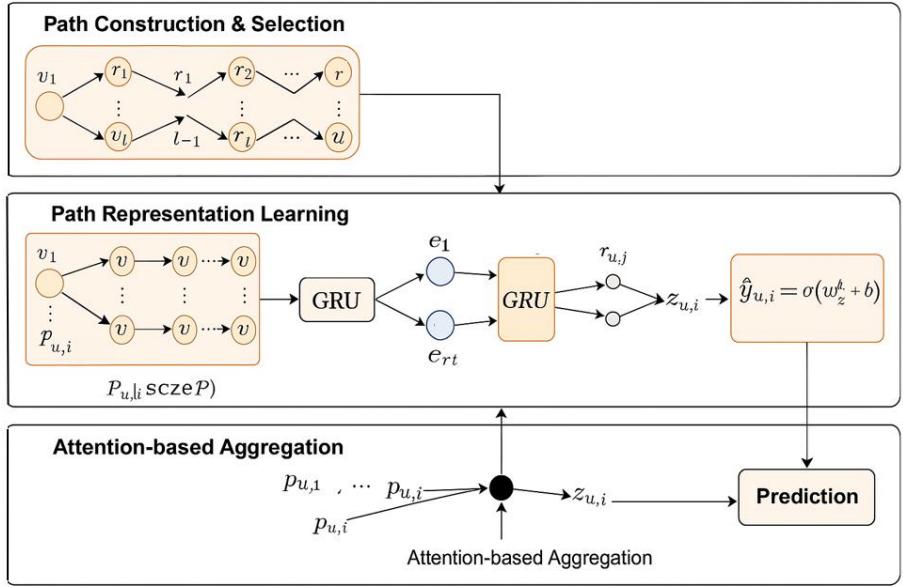

Figure 1. Overall model architecture

The model architecture, illustrated in Figure 1, delineates a three-stage pipeline designed to capture high-order semantics in heterogeneous information networks. The process begins with the construction of multi-hop paths linking users and items. To ensure that the input to the model remains both informative and efficient, a path screening module evaluates the contextual and structural quality of each path, retaining only those that exhibit strong semantic alignment. This approach leverages recent insights into dynamic rule filtering [9], which have proven effective in heterogeneous and noisy environments. Following path selection, the model employs a GRU-based encoder to transform each path into a dense vector representation. The GRU sequentially encodes the entities and relations along a path, preserving both short-term and long-range dependencies. This representation captures the semantic transitions across the path, ensuring that the resulting embeddings reflect both structure and meaning [10]. In the final stage, the model aggregates the path-level embeddings using an attention mechanism. This mechanism assigns relevance scores to each path, enabling the system to prioritize those most predictive of user preference. Recent probabilistic modeling approaches [11] support the use of such adaptive weighting to handle structural diversity and to improve robustness in downstream tasks. The resulting attention-weighted representation forms the basis of the model's final recommendation score.

First, a set of multi-hop paths from users to items is constructed, with the path length set to $l \in [2, L]$, where each path $P = (v_1, r_1, v_2, ..., r_{l-1}, v_l)$ is represented as a sequence of multiple entities and relationships. In order to reduce the interference of redundant paths and irrelevant information, a strategy based on path frequency and local mutual information is used to screen candidate paths to obtain a high-quality path subset $P_{ui} \subseteq P$, where $P_{ui}$ represents the set of all valid paths between user u and item i.

Next, in order to fully capture the high-order semantics of entities and relations in the path, a path encoder is used to learn the representation of each path [12]. Specifically, for the t-th node $v_t$ and relationship $r_t$ in the path, its embedding representation is $e_{v_t}, e_{r_t}$, and the entire path representation is modeled by GRU to retain the sequence dependency characteristics. The formula is as follows:

$$h_t = GRU(h_{t-1}, [e_{v_t}; e_{r_t}])$$

The final path is represented as the last hidden state $h_t$, which is $p_{u,i}^{(j)} = h_l$. In order to distinguish the importance of different paths to the recommendation results, the path attention mechanism is introduced to weight the different paths and aggregate them:

$$a_j = \frac{\exp(w^T \tanh(W_a p_{u,i}^{(j)}))}{\sum_k \exp(w^T \tanh(W_a p_{u,i}^{(k)}))}$$

Where $a_j$ represents the attention weight of path j and $w, W_a$ is a learnable parameter. The path-aware matching between the final user and the item is represented as the weighted sum of all paths:

$$z_{u,i} = \sum_j a_j \cdot p_{u,i}^{(j)}$$

Finally, the path-aware matching representation is input into the prediction module to generate the preference score $y_{u,i}$ of user u for item i, and trained with binary cross entropy loss:

$$y'_{u,i} = \sigma(w_z^T z_{u,i} + b)$$

$$L = -\sum_{(u,i) \in D} y_{u,i} \log(y'_{u,i}) + (1 - y_{u,i}) \log(1 - y'_{u,i})$$

$\sigma$ is the sigmoid function, $D$ is the training sample set, and $y_{u,i} \in \{0,1\}$ indicates whether the user interacts with the item. This method achieves unified modeling of path selection, semantic expression, and preference prediction through path-level modeling and attention mechanisms and improves the utilization efficiency of multi-hop structural information in heterogeneous information networks.

III. EXPERIMENT

A. Datasets

This study uses the Amazon-Book dataset as the primary source of experimental data. The dataset originates from the Amazon e-commerce platform and constructs a heterogeneous information network consisting of multiple node types and interaction relations, including users, items, categories, and brands. It is widely used for the validation and evaluation of heterogeneous recommendation systems. The dataset contains over 80,000 users, more than 200,000 items, and their corresponding category information. On average, each user interacts with more than ten items through actions such as purchases and ratings.

To support multi-hop path modeling, this study constructs several types of meta-paths based on the raw data, including "user-item-category," "user-item-brand," and "user-user-item." These meta-paths enhance the structural diversity and path expressiveness of the network. All nodes and relations are mapped into low-dimensional embedding vectors and normalized to ensure stable training and consistent representation.

In the experimental setup, users' historical interactions are divided into training and test sets in chronological order. The training set is used for multi-hop path sampling and representation learning, while the test set is used to evaluate recommendation performance. To address cold-start and data sparsity issues, users and items with very few interaction records are filtered out. As a result, a structurally rich, semantically clear, and path-aware modeling-ready heterogeneous recommendation dataset is constructed.

B. Experimental Results

First, this paper conducted a comparative test, and the experimental results are shown in Table 1.

Table 1. Comparative experimental results

| Method | HR@10 | RECALL @10 | PRECISION @10 |
|---|---|---|---|
| MF[13] | 0.6184 | 0.4325 | 0.3845 |
| NeuMF[14] | 0.6542 | 0.4578 | 0.4023 |
| GCN-Rec[15] | 0.6725 | 0.4691 | 0.4156 |
| HIN-PathRank[16] | 0.6893 | 0.4816 | 0.4238 |
| Ours | 0.7137 | 0.4982 | 0.4417 |

The experimental results show that the proposed path-aware recommendation method outperforms all baseline models across all evaluation metrics. In terms of HR@10, the model achieves a score of 0.7137, which is significantly higher than traditional methods such as MF and NeuMF. This indicates that the proposed method has a stronger ability to accurately hit users' true interest items within the recommendation list. In contrast, although GCN-Rec and HIN-PathRank incorporate structural information to some extent, they still fail to fully capture the high-order semantics of multi-hop paths in heterogeneous networks.

Further examining the RECALL@10 and PRECISION@10 metrics, the proposed method reaches 0.4982 and 0.4417, respectively, outperforming all baseline models. This suggests that the model not only retrieves more relevant items but also improves recommendation quality while effectively reducing false positives. These results indicate that the path-aware mechanism has clear advantages in selecting and representing high-quality paths, contributing to a more accurate depiction of user preferences.

In summary, the model enhances overall recommendation performance by integrating path selection, semantic modeling, and attention-based fusion strategies. Compared with models that rely only on adjacency information or predefined path structures, the proposed approach demonstrates greater adaptability and semantic expressiveness. This confirms the effectiveness and advancement of multi-hop path modeling in heterogeneous information network-based recommendation tasks.

Furthermore, this paper presents an experiment on the impact of different path lengths on recommendation performance, and the experimental results are shown in Figure 2.

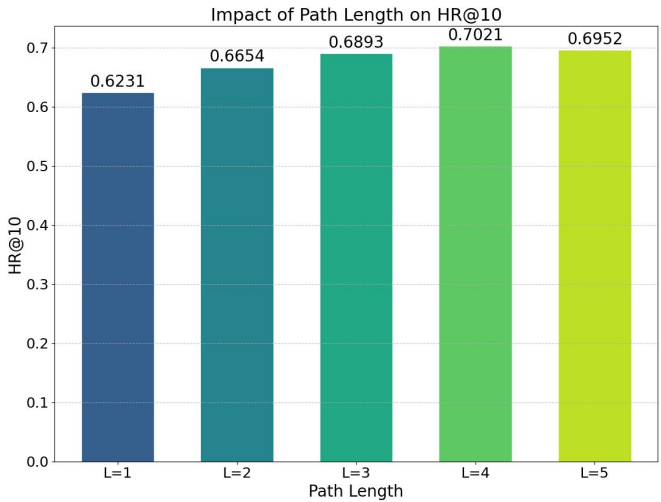

Figure 2. Experiment on the impact of different path lengths on recommendation performance

The results in the figure show that the increase in path length has a significant impact on recommendation performance. Short paths (e.g., L=1 and L=2) achieve relatively low HR@10 scores, with 0.6231 and 0.6654, respectively. This indicates that relying solely on shallow adjacency relations is insufficient to capture complex user interest structures.

As the path length increases to L=3 and L=4, model performance gradually improves, reaching the optimal value of 0.7021 at L=4. This demonstrates that multi-hop paths can effectively expand the semantic space of users. By traversing multiple types of nodes and relations, the model gains a stronger ability to represent deeper user preferences.

However, when the path length is further increased to L=5, the performance slightly declines to 0.6952. This suggests that overly long paths may introduce redundant or noisy information, which weakens the expressiveness of valid paths. Therefore, properly setting the path length is a key factor in improving recommendation quality. A balance must be achieved between representational power and semantic dilution.

Finally, the loss function drop graph is given, as shown in Figure 3.

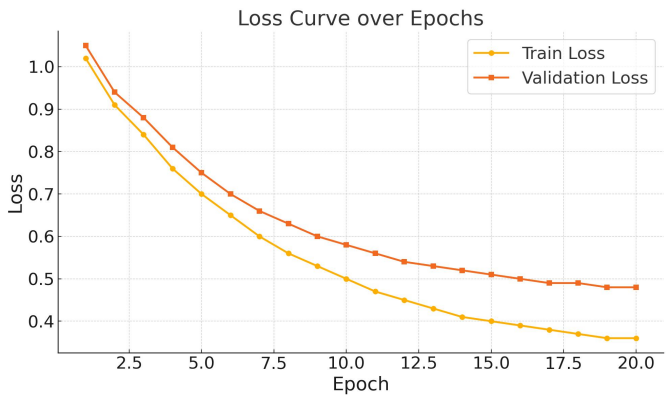

Figure 3. Loss function drop graph

From the loss function decline graph, we can see that with the increase of training rounds, the overall loss value of the model on the training set and the validation set shows a steady downward trend. This shows that the model is continuously learning the complex interaction pattern between users and items and gradually converging to a better state.

The training loss decreases slightly faster than the validation loss. The validation set decreases slowly after the 10th round but still maintains a stable downward trajectory. This shows that the model has good generalization ability during the training process, and there is no obvious overfitting phenomenon, which verifies the stability of the path-aware modeling mechanism in capturing high-order semantic relationships.

In addition, the final validation loss is stable at a low level, indicating that the proposed method can not only fully fit the training data but also has strong predictive ability on unseen data. This provides strong support for the reliability and practicality of subsequent recommendation tasks.

## IV. CONCLUSION

This paper proposes a multi-hop path-aware recommendation algorithm to address the problem of multi-hop path modeling in heterogeneous information networks. The method constructs high-quality multi-hop paths between users and items. By combining sequential modeling with attention mechanisms, it effectively captures high-order semantic information within paths and significantly improves the performance of the recommendation system. Experimental results show that the proposed model outperforms mainstream methods across multiple evaluation metrics, confirming the feasibility and effectiveness of path modeling in heterogeneous recommendation tasks. This study highlights the central role of multi-hop paths in representing complex user interest structures. By introducing path filtering and semantic-aware mechanisms, the model enhances its ability to express users' latent preferences. It also improves the modeling accuracy of multi-level factors such as behavioral transitions and semantic fusion. This modeling approach offers a new perspective beyond traditional shallow interaction-based recommendations and provides a practical solution for semantic mining in complex structured data.

From architectural design to module integration, this study presents a scalable path modeling framework that can be applied to various types of heterogeneous network structures. The method is suitable not only for e-commerce recommendation but also has potential applications in social networks, content delivery platforms, and other domains. Meanwhile, the performance of the path-aware mechanism in handling high-order structural relations and filtering redundant information further expands the system's capability to understand multi-source heterogeneous data. Future work may proceed in two directions. First, exploring more adaptive path generation strategies to support dynamic modeling in online recommendation scenarios [17-18]. Second, integrating advanced methods such as graph neural networks and contrastive learning to further optimize path representation learning and semantic discrimination [19]. This study provides a theoretical foundation and practical exploration for structured path modeling in heterogeneous network recommendation, offering both academic value and engineering potential.